%% file: main.tex
\documentclass{article} 
\usepackage{iclr2021_conference,times}

\input{math_commands.tex}

\usepackage{hyperref}
\usepackage{url}

\usepackage{booktabs}
\usepackage{multirow}
\usepackage{siunitx}
\usepackage{graphicx}

\AtBeginDocument{%
  \providecommand\BibTeX{{%
    \normalfont B\kern-0.5em{\scshape i\kern-0.25em b}\kern-0.8em\TeX}}}

\newif\iffinal
\finaltrue

\iffinal
  \newcommand\sutanay[1]{}
  \newcommand\logan[1]{}
  \newcommand\jenna[1]{}
  \newcommand\ganesh[1]{}
  \newcommand\neeraj[1]{}
  \newcommand\ian[1]{}
\else
  \newcommand\sutanay[1]{{\color{blue}[Sutanay: #1]}}
  \newcommand\logan[1]{{\color{violet}[Logan: #1]}}
  \newcommand\jenna[1]{{\color{green}[Jenna: #1]}}
  \newcommand\ganesh[1]{{\color{red}[Ganesh: #1]}}
  \newcommand\neeraj[1]{{\color{cyan}[Neeraj: #1]}}
  \newcommand\ian[1]{{\color{orange}[Ian: #1]}}
\fi

\newcommand\icfifty{IC\textsubscript{50}}

\begin{document}







\title{Evening the Score: Targeting SARS-CoV-2 Protease Inhibition in Graph Generative Models for Therapeutic Candidates}

\author{Jenna Bilbrey, Sutanay Choudhury, Neeraj Kumar\\
Pacific Northtwest National Laboratory\\
Richland, WA, USA \\
\texttt{\{jenna.pope, sutanay.choudhury, neeraj.kumar\}@pnnl.gov} \\
\AND
Logan Ward, Ganesh Sivaraman\\
Argonne National Laboratory\\
Lemont, IL, USA \\
\texttt{\{lward, gsivaraman\}@anl.gov} \\
}


\iclrfinalcopy
\maketitle
\begin{abstract}
We examine a pair of graph generative models for the therapeutic design of novel drug candidates targeting SARS-CoV-2 viral proteins. Due to a sense of urgency, we chose well-validated models with unique strengths: an autoencoder that generates molecules with similar structures to a dataset of drugs with anti-SARS activity and a reinforcement learning algorithm that generates highly novel molecules. During generation, we explore optimization toward several design targets to balance druglikeness, synthetic accessability, and anti-SARS activity based on \icfifty. This generative framework\footnote{https://github.com/exalearn/covid-drug-design} will accelerate drug discovery in future pandemics through the high-throughput generation of targeted therapeutic candidates. 
\end{abstract}

\section{Introduction}

\begin{figure}[!h] \centering
  \includegraphics[width=0.95\textwidth]{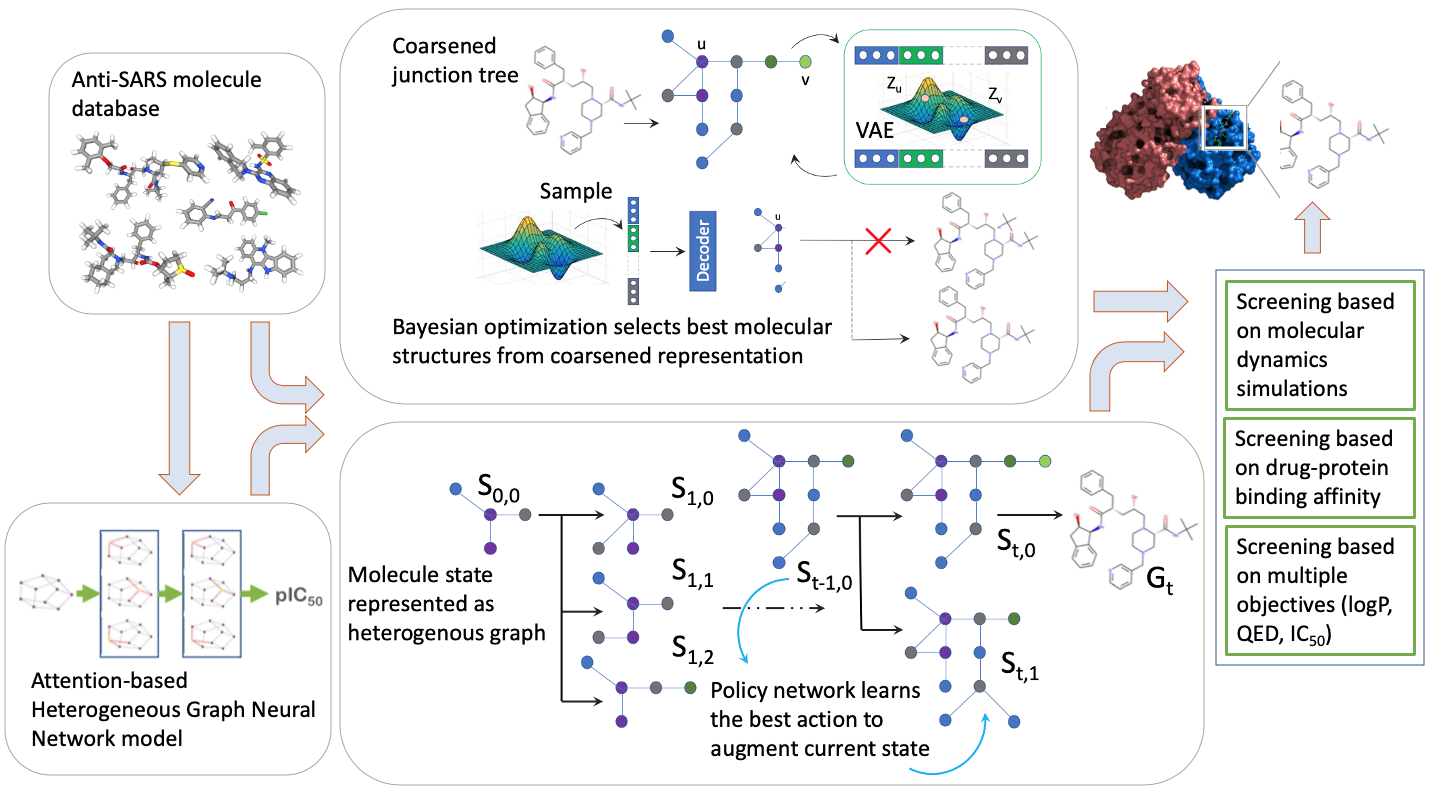}
  \caption{Depiction of workflow developed for this work. The anti-SARS database is used to train a MPNN to predict pIC\textsubscript{50} and the JT-VAE model. The trained MPNN is used as the scoring function in both JT-VAE (top row) and DQN-based molecular generation (bottom). Candidate molecules are screened by pIC\textsubscript{50} ($>$8) and validated by a Drug Target Binding Affinity classifier.} \label{fig:mol_design_architecture}
\end{figure}

Discovering lead candidates for SARS-CoV-2 presents a challenge to the scientific community, and systematically developing AI-based automated workflows can accelerate the design of effective therapeutic compounds with a target set of properties [\cite{GCPN, Li2018GraphGen, huang2020deeppurpose, COVID-repurpose-nature, horwood2020molecular, khemchandani2020deepgraphmol, IBM, hu2020generating, selfiegeneration2020, joshi2021artificial}]. The potential chemical space is composed of over $10^{60}$ compounds that ideally need to be tested for the therapeutic design [\cite{reymond2015chemical}] and discovery before identifying a lead for a given target protein of SARS-CoV-2. However, candidates with suitable activity against specific proteins only narrows the search space to $10^4-10^5$  compounds. There is a need for rapid development of machine-learning (ML) workflows that serve as a search-and-screen process of this reduced chemical space. 
Candidate molecules generated by ML models are passed to downstream verification via virtual high-throughput drug-protein binding techniques, chemical synthesis, biochemical and biophysical assay, and finally clinical trials [\cite{Batra2020Covid, SummitDocking}]. 
In the current work, we develop, evaluate, and compare two leading graph-generative approaches to design candidate therapeutic compounds targeting SARS-CoV-2 protease inhibition (Figure \ref{fig:mol_design_architecture}). This systematic comparison of two leading approaches distinguishes our work in the context of the broad effort undertaken by the research community in its quest for novel therapeutics (see section \ref{sec:related_work} for a thorough review).

First, we train a junction-tree variational autoencoder (JT-VAE) [\cite{JTVAE}] to generate molecules with similar structures to a curated dataset of known drugs with anti-SARS activity. The trained JT-VAE is then used in Bayesian optimization to generate novel candidates with targeted properties. We next examine graph-based deep reinforcement learning (DQN) [\cite{MolDQN}] to generate candidates that are not constrained by their structural proximity to known anti-SARS compounds. To benchmark the two graph generative models, we use the same set of property scoring functions as optimization targets. 

\textbf{Contributions.} 
The goal of our study is to perform multi-objective optimization to generate molecular candidates by considering important bioactivity properties along with druglikeness and synthetic accessability. We focus on pIC\textsubscript{50} (the inverse log of the half maximal inhibitory concentration, IC\textsubscript{50}), which is an experimentally measured property that captures the potency of a therapeutic candidate towards a protease target, where higher values indicate exponentially more potent inhibitors. Notably, p\icfifty~cannot be accurately modeled through \textit{ab initio} methods [\cite{IC50-QM, sebaugh2011guidelines}]. 
Our ML workflow for the high-throughput generation of therapeutic candidates with anti-SARS activity contains the following key components: a surrogate model for pIC\textsubscript{50} prediction, candidate generation via JT-VAE and DQN models, and validation of top-ranking candidates against a Drug Target Binding Affinity (DBTA) classifier [\cite{huang2020deeppurpose}] to asses their potential activity against SARS-CoV-2. Alongside this paper, we release the ML workflow and a curated dataset of drug molecules with anti-SARS activity for use by the community. 

\section{Background and Related Work}
\label{sec:related_work}
\textbf{Biological factors in drug design.} A variety of properties can be considered for lead optimization [\cite{copeland2006drug}]. Here we give an explanation of the various properties and their ideal values. Lipinski's Rule of Five is the standard starting place when considering properties of potential oral drug compounds [\cite{lipinski2004lead}]. The Rule of Five identifies that molecules with no more than 5 hydrogen bond donors, no more than 10 hydrogen bond acceptors, a molecular mass under 500 Da, and an octanol--water partition coefficient (log P) below 5 are more likely to show pharmacological activity. LogP is a measure of lipophilicity, which provides an understanding of the behavior of a drug in the body. More recent analysis has shown that logP values between 1 and 3 may be more appropriate considering the effect of logP on absorption, distribution, metabolism, elimination, and toxicology (ADMET) properties [\cite{logP2010}]. Though oral bioavailability is an important factor, a sole focus on logP has the potential to screen out otherwise useful compounds [\cite{Beyond5}]. QED has been proposed as a more holistic druglikeness metric [\cite{QED}], which from 0 (low) to 1 (high). However, druglikeness does not indicate the activity or effectiveness of a drug towards a specific target. The half maximal inhibitory concentration (IC\textsubscript{50}) provides a quantitative measure of the potency of a compound to inhibit a specific biological process. IC\textsubscript{50} is obtained by measurement, and no universal \textit{ab initio} method of computing its value exists. A number of methods have been developed to approximate IC\textsubscript{50}, many based on QSPR and recently some based on machine learning [\cite{IC50-QSAR, IC50-QM, IC50-ML, IC50-LSTM}]. Similarity to known drugs is also an important factor in drug discovery [\cite{Similarity2014, Similarity2018}], as is the ability to synthesize the molecule, which can be estimated by the synthetic accessibility (SA) score -- from 1 (easy) to 10 (difficult) [\cite{SAScore}].

\textbf{COVID-19-focused efforts.} In response to the COVID-19 pandemic, researchers have pushed to identify marketed drugs that can be repurposed for SARS-CoV-2 treatment [\cite{COVID-repurpose-nature, SummitDocking, CovidInteractionMap, zhavoronkov2020potential, skwark2020designing, mittal2020identification, Indinavirdocking, chen2020prediction, wu2020analysis}]. Most of these the COVID-19-focused efforts relied on molecular simulations and docking studies to identify known drugs that interacted with 3-chymotrypsin-like main protease (3CLpro), also called the main protease (Mpro), of SARS-Cov-2. Machine learning (ML) approaches to drug repurposing were also reported. For instance, \cite{Batra2020Covid} applied an ML-based screening approach to find known compounds with binding affinity to either the isolated SARS-CoV-2 S-protein at its host receptor region or to the S-protein-human ACE2 interface complex. \cite{huang2020deeppurpose} developed DeepPurpose, a deep learning toolkit for drug repurposing, with the goal of recommending known candidates with high binding affinity to target amino acid sequences.

An array of \textit{de novo} design approaches specifically targeting anti-SARS-CoV-2 drugs have also been reported. \cite{hu2020generating} trained the GPT2 transformer-based model to generate SMILES strings of potential drug candidates follow the Lipinski's "Rule of Five", optimizing towards 3CLpro--target interaction. \cite{IBM} developed the generative modeling framework CogMol which relies upon a varitional autoencoder sampled by Controlled Latent attribute Space Sampling (CLaSS) to design candidates optimized toward binding affinity. \cite{bornpaccmannrl} amended their PaccMann RL approach, which coupled reinforcement learning with variational autoencoders with reinforcement learning, to generate molecules optimized towards binding affinity and pharmacological toxicity predictors. \cite{zhavoronkov2020potential} applied a similar approach of using reinforcement learning to optimize a generative model, but examined genetic algorithms, language models, generative adversarial networks as well as autoencoders. \cite{tang2020ai} applied a deep Q learning approach with a structure-based optimization policy that targets QED, the presence of specific fragments, and the presence of pharmacophores. In a novel direction, \cite{skwark2020designing} focused on the design of proteins that bind to SARS-CoV-2 in place of ACE2 using reinforcement learning.

\textbf{Graph generative methods for molecule design.} Heterogeneous graphs provide a natural representation for small-molecule organic compounds, with nodes representing atoms in the molecular structure and edges representing bonds between the atoms [\cite{weininger1990smiles}]. This approach motivated the exploration of graph-generative models, such as graph convolutional policy networks [\cite{GCPN}], variational autoencoders [\cite{JTVAE, samanta2019nevae, JTVAE-multi}], and variants of deep reinforcement learning [\cite{MolDQN, staahl2019deep}], for the target-driven optimization of drug molecules.
In the current work, we develop, evaluate, and compare two leading graph-generative approaches for candidate therapeutic compounds, variational autoencoders and reinforcement learning, specifically targeting molecules able to inhibit the function of SARS-CoV-2 proteins. Our focus is to understand the distinction between the variational autoencoder and reinforcement learning approaches in terms of similarity versus novelty in the generated molecules, as well as the optimal approach to developing a wholistic scoring function that accounts for both druglikeness and SARS-Cov-2 protease inhibition.

\section{Method}


\subsection{Surrogate Model for \texorpdfstring{p\icfifty}~ Prediction}
\label{sec:mpnn}
We trained a message-passing neural network (MPNN) [\cite{scarselli2008graph, gilmer2017neural}] to predict pIC\textsubscript{50} (the inverse log of IC\textsubscript{50}) for a given molecular structure.
Following the formalism of \cite{gilmer2017neural}, our network is composed of message, update, and readout operations (eqns.\ 1-3) and our choices
for these functions are based on networks developed by \cite{stjohn2019mpnnpolymer} for polymer property prediction.

The original state of each atom ($h_v$) and bond ($\alpha_{vw}$) in our molecule ($G$) is a 256-length vector with values defined by an embedding table based on the atomic number and bond type (e.g., single, double, aromatic).
The states of these atoms are modified by eight successive ``message'' layers.
Each message layer uses a two-layer multi-layer perceptron (MLP) with sigmoid activations to compute a message that uses the state of an 
atom ($h_v$), the state of the neighboring atom ($h_w$) and the bond
which joins them ($\alpha_{vw}$). 
The atom and bond states are updated according to the following equations:

\begin{equation}
m^{t+1}_v = \sum_{w \in Neighbors(v)} M_t(h^t_v, h^t_w, \alpha^{t}_{vw})
\end{equation}
\begin{equation}
h^{t+1}_v = h^t_v + m^{t+1}_v
\end{equation}
\begin{equation}
\alpha^{t+1}_{vw} = \alpha^t_{vw} + M_t(h^t_v, h^t_w, \alpha^{t}_{vw})
\end{equation}

The atom states output from the last layer ($h^T_v$) are used to predict the p\icfifty~of the molecule using a "readout" function ($R$):

\begin{equation}
\hat{y}=R(h^T_v |v \in G)
\label{eqn:ic50_aggr}
\end{equation}

We examined several variants of the readout function in our study.
We tested both ``molecular fingerprints,'' where the states of each node are combined \textit{before} using a multi-layer perceptron (MLP) to reduce to compute pIC\textsubscript{50}, and an ``atomic contribution,'' where we combine the nodes \textit{after} MLP to compute a per-node contribution to pIC\textsubscript{50}.
We experimented with the use of five different functions to reduce the atomic state/contributions to a single value for each graph: summation, mean, maximum, softmax, and attention.
The attention functions are created by learning an attention map by passing the node states through a MPNN.
We tested all combinations of ``molecular fingerprint'' vs.\ ``atomic contribution'' and the five readout functions, for a total of 10 networks, training each on network the same 90\% of our  pIC\textsubscript{50} dataset and comparing its performance on the withheld 10\% of the data.
We used an MPNN that uses attention maps to reduce contributions from each atom to a 
single pIC\textsubscript{50} of a molecule in all subsequent experiments.



\textbf{Junction-Tree Variational Autoencoder.} We used a junction-tree variational autoencoder (JT-VAE) [\cite{JTVAE}] to generate molecules with high proximity to anti-SARS therapeutic molecules. The autoencoder generates novel molecular graphs by laying a tree-structured scaffold over substructures in the molecule, which are then combined into a valid molecule using a MPNN. JT-VAE allows for the expansion of a molecular graph through the addition of subgraphs, or ``vocabulary" of valid components, derived from the training set (Fig.~\ref{fig:mol_design_architecture}). The subgraphs are used to encode a molecule into its vector representation and decode latent vectors into valid molecular graphs. The use of subgraphs maintains chemical validity at each step, while also incorporating chemical units common to the training set. Chemical graphs generated from the vocabulary are structurally similar to those in the training set, which is a benefit when attempting to design molecules with similar properties to known drugs. We omit the details here for brevity and point the reader to \cite{JTVAE} for details on JT construction and \cite{GrammarBO} for Bayesian optimization. 

\textbf{Deep Reinforcement Learning.} We follow the Q-learning approach of Zhou et al.\ for our deep reinforcement learning approach [\cite{MolDQN}]. In this setting, we cast the molecule generation problem as a Markov Decision Process (MDP) [\cite{DQN}] to learn a policy network $\pi$ that determines the best sequence of actions to transform an initial molecular graph to a larger graph with desirable properties in a step-by-step fashion (Fig.~\ref{fig:mol_design_architecture}).  At each step, we enumerate all possible actions and select those which produce valid molecules (e.g., respect valency rules).  Next, we train a MLP to predict the \textsl{value} of a certain action by passing the Morgan fingerprints [\cite{rogers2010extended}] as input.  The MLP approximates the value of an action computed using the Bellman equation, where the score of a state and the maximum score of the subsequent state is multiplied by a decay factor.  As established with other Deep Q-Learning approaches, the addition of the value of the next state increases the value of moves which will lead to higher future rewards.


\textbf{SARS-CoV-2-specific Scoring Functions.} The scoring functions described in this section are used for both Bayesian optimization in the JT-VAE approach and reward computation for the deep reinforcement-learning approach. 

\begin{equation}
logP^P(m) = logP(m) - SA(m) - cycle(m)
\end{equation}
\begin{equation}
QED^P(m) = QED(m) - SA(m) - cycle(m)
\end{equation}
\begin{equation}
pIC_{50}(m) = MPNN(m)
\end{equation}
\begin{equation}
pIC_{50}+QED^P(m) = MPNN(m) + QED(m) - SA(m) - cycle(m)
\end{equation}

Following \cite{JTVAE}, we first compute a score that penalizes the partition coefficient between octanol and water (logP) by the synthetic accessibility (SA) score (in which higher SA values are discouraged) and the number of cycles with more than 6 atoms (eqn.~5). 
Considering that the Quantitative Estimate of Druglikeness (QED) is a more comprehensive heuristic than logP, we also use a similar scoring function composed of QED penalized by the SA score and number of long cycles (eqn.~6). We then examine the utility of a SARS-specific scoring function based on the pIC\textsubscript{50} predicted by our MPNN (eqn.~7). Finally, we examine a multi-objective scoring function that includes both pIC\textsubscript{50} and penalized QED (eqn.~8).

\section{Experimental Analysis}

The focus of our study is two fold: \textbf{1)} Perform multi-objective optimization for generating therapeutic candidates with activity towards protease target of SARS-CoV-2, \textbf{2)} Build novel yet unique compounds by optimizing the scoring function including synthesizibility score. To this end, we are poised to utilize best possible graph based ML models.



\label{sec:Dataset Preparation}

\textbf{Dataset Preparation.} 
We assembled and curated a protease dataset containing molecules active against various protease in enzymatic assays from experimental pharmacology databases, such as CheMBL, BindingDB, and ToxCat [\cite{bento2014chembl}]. The dataset was filtered based on pIC\textsubscript{50} activity and potency. Molecules larger than 1,000 Dalton were removed along with non-drug like molecules containing metals and polypeptides. 
In some instances, duplicate entries were generated from multiple experimental studies, in which case we used the mean pIC\textsubscript{50}. The resulting dataset contains SMILES strings and experimental pIC\textsubscript{50} values of 6,545 unique molecules.   

\textbf{Quantitative Characterization.} We computed the metrics included in our scoring functions for each molecule in the dataset. 
The highest pIC\textsubscript{50} is 10.89, while the lowest is 1.22. The most common pIC\textsubscript{50} is 4.0, which is shared by 320 structures, and the vast majority of structures (91.5\%) have pIC\textsubscript{50} values greater than 4.0. LogP values range from -10.36 to 16.65 in a near Gaussian distribution with a mean of 3.70; 77.0\% of all structures meet the requirement of Lipinski's Rule of 5 that logP be no greater than 5. QED values range from 0.01 to 0.94, while SA values range from 1.35 to 8.24, with 73.8\% being below 4. We computed the Tanimoto similarity [\cite{Tanimoto2015}] for all pairs of compounds to gain insight into the structural diversity of molecules in our dataset (Fig.~\ref{fig:database_similarity}). 
We observed that structures tend to become more similar to their neighbors as pIC\textsubscript{50} increases, indicating that compounds with high pIC\textsubscript{50} values tend to be structurally similar, supporting the consideration of molecular similarity during drug design and discovery.

\textbf{Training pIC\textsubscript{50} Surrogate Model.} 
We found that limiting the pIC\textsubscript{50} prediction to contributions from only a few specific atoms in the molecule improved performance. This improved performance can be explained by a physical mechanism. The presence or absence of a specific chemical pattern in a molecular structure (e.g., functional groups, substituents, aromatic rings) controls binding of the molecule to a certain portion of a target protein, while the other atoms in the molecule play a role in determining whether the molecule will stay affixed at the target site. We hypothesize that such contributions can have a critical effect in tuning the binding affinity, which influences p\icfifty. 



\textbf{JT-VAE Setup.} We trained the JT-VAE on our dataset for 8,300 iterations with the following hyperparameters: hidden state dimension of 450, latent code dimension of 56, and graph message passing depth of 3. 
To optimize towards the specified scoring functions, we trained a sparse Gaussian process (SGP) to predict a score given the latent representation learned by the JT-VAE and then performed 10 iterations of batched Bayesian optimization (sampling = 50) based on the expected improvement. 

\logan{I moved this from later in this section to make a focused section on comparing the performance.}

\begin{table}
\caption{Properties of the top-3 molecules generated using the specified scoring function. 
}
  \resizebox{\textwidth}{!}{\begin{tabular}{lSSSSSSSSSSSSSSS}
    \toprule
    \multirow{2}{*}{Scoring Function } &
      \multicolumn{3}{c}{pIC\textsubscript{50}} &
      \multicolumn{3}{c}{QED} &
      \multicolumn{3}{c}{logP} & 
      \multicolumn{3}{c}{SA Score}\\
       & {1st} & {2nd} & {3rd} & {1st} & {2nd} & {3rd} & {1st} & {2nd} & {3rd}& {1st} & {2nd} & {3rd}\\
      \midrule
    logP$^P$ (JT-VAE)	&	4.93	&	4.57	&	4.60	&	0.45	&	0.78	&	0.71	&	4.05	&	4.27	&	4.20	&	1.70	&	2.08	&	2.08	\\
    logP$^P$ (DQN)  &  6.10 &  8.17 &  4.98 &  0.04 &  0.07 &  0.11 &  13.86 &  12.64 &  12.52 & 3.59 & 2.97 & 2.93 \\
    \hline
    QED$^P$	 (JT-VAE)&	4.15	&	4.23	&	4.71	&	0.91	&	0.84	&	0.91	&	3.72	&	2.19	&	2.20	&	1.80	&	1.71	&	2.25	\\
    QED$^P$	(DQN) &  6.80 &  6.80 &  6.80 &  0.77 &  0.77 &  0.77 &  3.46 &  3.46 &  3.46 & 2.05 & 2.05 & 2.05	\\
    \hline
    pIC\textsubscript{50} (JT-VAE)	&	10.22	&	10.07	&	10.07	&	0.15	&	0.12	&	0.12	&	4.86	&	3.80	&	3.06	&	4.52	&	4.94	&	4.98	\\
    pIC\textsubscript{50} (DQN) &  10.57 &  10.56 &  10.39 &  0.09 &  0.11 &  0.41 &  1.51 &  2.03 & -0.44 & 6.90 & 6.71 & 5.57 \\
    \hline
    pIC\textsubscript{50}+QED$^P$ (JT-VAE)  &	8.58	&	5.98	&	8.18	&	0.83	&	0.87	&	0.70	&	4.02	&	3.37	&	3.82	&	1.93	&	1.74	&	1.50\\
    pIC\textsubscript{50}+QED$^P$ (DQN)   &  10.27 &  10.27 &  10.27 &  0.80 &  0.80 &  0.80 &  3.02 &  3.02 &  3.02 & 2.90 & 2.90 & 2.90 \\
    \bottomrule
      \end{tabular}}\label{tab:jtvae}
\end{table}

\textbf{DQN Setup.} 
Our DQN approach constructs molecules atom-by-atom and bond-by-bond in a step-wise manner.
Each episode starts with either a single atom or a pre-seeded molecule and the RL agent is allowed up to 40 steps to construct an optimized molecule.
A move is determined either randomly or by the Q-function model.
We update the model to predict the move after each step in each episode and, as this model improves during training, we gradually reduce the probability that a random move is chosen over the model prediction. 
If we start with a single atom approach the DQN finds tens of thousands of candidate molecules that needed to be optimized. 
Starting from a pre-seeded molecule led to an increased number of molecules generated with high p\icfifty, while starting from a single atom produced molecules with higher QED, as shown in Fig.~\ref{fig:pareto_plot}.

\logan{Be consistent about ``scoring'' function and not reward function.}


\textbf{Comparative Effect of the Scoring Function.} 
Table \ref{tab:jtvae} shows the top-3 molecules generated by the two generative models with each scoring function. 
DQN always outperforms JT-VAE in finding a molecule with a superior value of the scoring function being optimized.
The performance disparity is particularly apparent when optimizing for logP: the maximum logP from DQN is 12.6 compared to only 4.1 for JT-VAE.
We attribute the difference in performance to JT-VAE implicitly sampling from a distribution of drug-like molecules which typically have logP values between -0.4 and 5.6, while DQN has no such constraints.
However, the JT-VAE-generated molecules have consistently better QED and SA scores even when those values are not explicitly optimized for. 
The molecules in our dataset were experimentally synthesized, leading the SA scores of these molecules to be quite low. Likewise, because they are drug molecules, their QED are high.
This leads JT-VAE to generate similar molecules with high QED and low SA scores.
On the other hand, the RL agent uses no information about the space of experimentally studied drug candidates during its training process and, accordingly, finds molecules far outside that region.

Overall, we find different purposes for JT-VAE and RL-based molecular optimization.
JT-VAE implicitly uses the distribution of molecules in its training set to 
bias towards realistic molecules, albeit at the expense of highly novel candidates.
The RL-based approach lacks such constraints and, for better or worse, optimizes without implicitly regarding synthesizability or other characteristics not explicitly encoded in the scoring function,  leading to many more novel candidates which are otherwise impossible to optimize toward.

\begin{figure}[htbp!] \centering
  \includegraphics[clip, trim=2.7cm 6.2cm 2cm 6.2cm, width=0.9\textwidth]{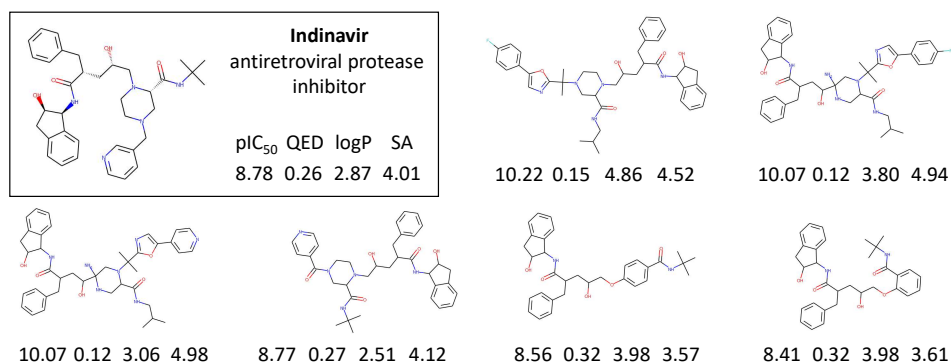}
  \caption{Top-6 molecules generated by the JT-VAE method optimized towards pIC\textsubscript{50}. 
  }
  \label{fig:topmolecules}
\end{figure}

\textbf{Qualitative Analysis.} We observed an interesting structural trend in the JT-VAE-generated molecules optimized towards pIC\textsubscript{50}. Fig.~\ref{fig:topmolecules} shows the structures of generated molecules with pIC\textsubscript{50} $>$ 8 and the anti-HIV drug Indinavir. A common backbone is shared between Indinavir and the top-6 predictions. The Tanimoto similarity of these six generated molecules against Indinavir range from 0.65 to 0.91. Indinavir has been proposed as a drug candidate to treat SARS-CoV-2 due to favorable binding affinity to the 3-chymotrypsin-like main protease (3CLpro) that is pivotal for the replication of SARS-CoV-2 [\cite{yoshino2020identification, Indinavirdocking, 3CLforCOVID,harrison2020coronavirus}].
Notably, three of the generated molecules have a higher predicted pIC\textsubscript{50} than Indinavir. 


\emph{In silico} Drug-Target Binding Affinity (DTBA) methods offer a way to evaluate drug--target interactions [\cite{he2017simboost}]. We employ a ML-based DTBA model to validate the interaction of molecules generated by JT-VAE against 3CLpro [\cite{chen2020prediction}]. We trained a DBTA binary classification model using the extended connectivity fingerprint [\cite{rogers2010extended}] encoding for the drug molecule and the target protease sequence encoding from the DeepPurpose toolkit [\cite{huang2020deeppurpose}]. 
The DBTA model classified 4 of the top-11 molecules (including the top-2 in Fig.~\ref{fig:topmolecules}) with $>$50\% probability of interaction with 3CLpro.

\section{Conclusions}
To possibly explore the novelty and uniqueness of generated molecules, we compared two graph generative models, JT-VAE and DQN, for the task of generating small-molecule therapeutic candidates with activity against 3CLpro target of SARS-CoV-2. DQN always outperformed JT-VAE in finding molecules with a superior value of the scoring function being optimized. However, JT-VAE generated molecules that were more structurally similar to those in the dataset due to substructure representation, which produced a lower SA score and logP $<$ 5. 


\bibliographystyle{iclr2021_conference}
\bibliography{citations}


\input{appendix}
\end{document}

%% file: math_commands.tex
\usepackage{amsmath,amsfonts,bm}

\def\eqref#1{equation~\ref{#1}}

\def\1{\bm{1}}

\DeclareMathAlphabet{\mathsfit}{\encodingdefault}{\sfdefault}{m}{sl}
\SetMathAlphabet{\mathsfit}{bold}{\encodingdefault}{\sfdefault}{bx}{n}



%% file: appendix.tex
\renewcommand{\thefigure}{S\arabic{figure}}
\setcounter{figure}{0} 
\renewcommand{\thetable}{S\arabic{table}}
\setcounter{table}{0} 
\setcounter{equation}{0}
\renewcommand{\theequation}{S\arabic{equation}}
\setcounter{page}{1}
\renewcommand{\thepage}{S\arabic{page}}

\clearpage

\section*{Supplementary Material}

\appendix

\section{Dataset Analysis}
\begin{figure*}[htbp!] \centering
  \includegraphics[clip, trim=0.5cm 6.5cm 0.15cm 6.5cm, width=0.9\textwidth]{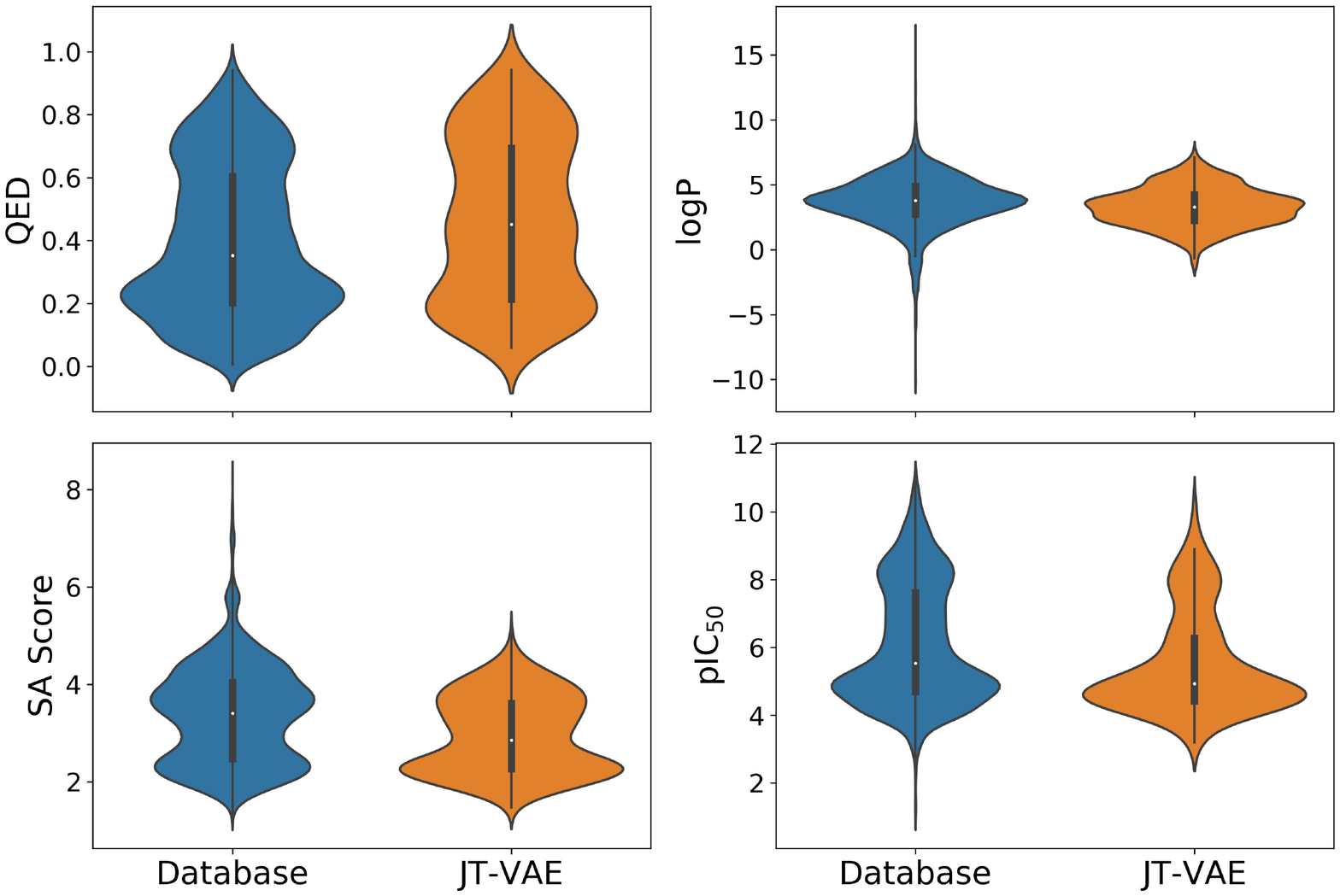}
  \caption{We generated 1,000 molecules using the trained JT-VAE, of which 560 were unique. The figure shows the comparison of the QED, logP, SA score, and pIC\textsubscript{50} of compounds in the database and those generated by the JT-VAE. The JT-VAE reproduced the range of values present in the database, minus outliers. The similar values indicate that the JT-VAE is able to reproduce the wide range of structures present in the database. The pIC\textsubscript{50} values for generated molecules were estimated by our MPNN.} \label{fig:DBvVAE}
 \end{figure*}

We computed the metrics included in our scoring functions for each molecule in the database. The range of values can be seen in Figure \ref{fig:DBvVAE}. The highest pIC\textsubscript{50} is 10.89, while the lowest pIC\textsubscript{50} is 1.22. The most common pIC\textsubscript{50} is 4.0, which is shared by 320 structures, and the vast majority of structures (91.5\%) have pIC\textsubscript{50} values greater than 4.0. LogP values range from -10.36 to 16.65 in a near Gaussian distribution with a mean of 3.70; 77.0\% of all structures meet the requirement of Lipinski's Rule of 5 that logP be no greater than 5. QED values range from 0.01 to 0.94, while SA values range from 1.35 to 8.24, with 96.3\% being below 5.

Figure \ref{fig:DBvVAE} also shows the range of properties of molecules generated by our trained JT-VAE without optimization. In general, JT-VAE reproduced the distribution of logP, QED, p\icfifty, and SA scores present in the database. The p\icfifty~values of the JT-VAE-generated molecules were obtained from our trained MPNN, discussed in Section \ref{sec:mpnn}.

\begin{figure*}[htbp!] \centering
  \includegraphics[clip, trim=0.15cm 5.2cm 1.5cm 5.2cm, width=0.8\textwidth]{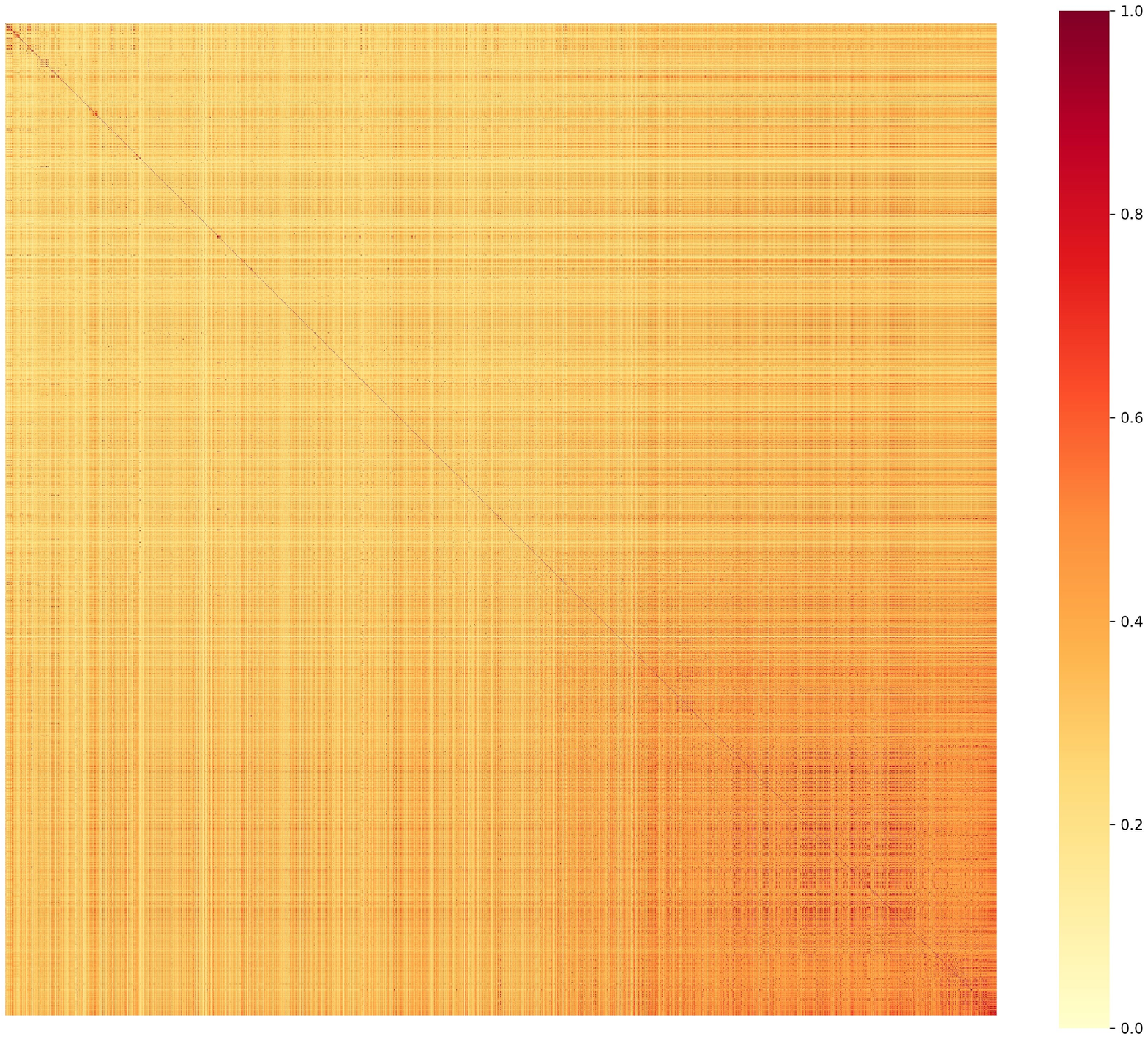}
  \caption{Heat map of the Tanimoto similarity between all compounds in the database. Entries are ordered in increasing pIC\textsubscript{50}. The Tanimoto similarity is generally higher among structures with high pIC\textsubscript{50} (located towards bottom right of the matrix), indicating the importance of considering structural similarity in drug discovery.} \label{fig:database_similarity}
 \end{figure*}

We also computed the Tanimoto similarity for all pairs of compounds to gain insight into the structural diversity of molecules in our database (Figure \ref{fig:database_similarity}). The entries in the matrix were ordered in increasing pIC\textsubscript{50} values.  The similarity is represented by the color bar, with yellow representing low similarity (0) and red high similarity (1). We observe that structures tend to become more similar to their neighbors as pIC\textsubscript{50} increases, indicating that compounds with high pIC\textsubscript{50} values tend to be structurally similar, supporting the consideration of molecular similarity during drug discovery.

\clearpage
\section{Effects of Pre-seeding RL}

\begin{figure}[htbp!] \centering
  \includegraphics[width=0.5\textwidth]{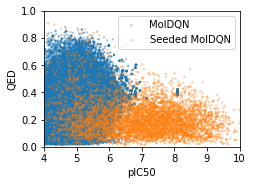}
\caption{(QED vs p\icfifty~for molecules generated by RL with starting from a single atom (MolDQN) or pre-seeded with a drug-like molecule (Seeded MolDQN) optimized towards the multi-objective scoring function in eqn.~S8.
}
\label{fig:pareto_plot}
\end{figure}

\clearpage